\documentclass[aps,preprint,eqsecnum,nofootinbib]{revtex4}

\usepackage{epsfig}
\usepackage[usenames, dvipsnames]{color}

\newcommand{\be}{\begin{equation}}
\newcommand{\ee}{\end{equation}}
\newcommand{\bea}{\begin{eqnarray}}
\newcommand{\beas}{\begin{eqnarray*}}
\newcommand{\eea}{\end{eqnarray}}
\newcommand{\eeas}{\end{eqnarray*}}
\newcommand{\ba}{\begin{array}}
\newcommand{\ea}{\end{array}}

\def\ls{\mathrel{\lower4pt\vbox{\lineskip=0pt\baselineskip=0pt
           \hbox{$<$}\hbox{$\sim$}}}}
\def\gs{\mathrel{\lower4pt\vbox{\lineskip=0pt\baselineskip=0pt
           \hbox{$>$}\hbox{$\sim$}}}}

\begin{document}

\title{Generating $\theta_{13}$ from sterile neutrinos in 
$\mu - \tau$ symmetric models}

\author{Diana C. Rivera-Agudelo~$^1$\footnote{email: drivera@fis.cinvestav.mx},
 Abdel P\'erez-Lorenzana~$^{1,2}$\footnote{email: aplorenz@fis.cinvestav.mx}}

\affiliation{
$^1$Departamento de F\'{\i}sica, Centro de Investigaci\'on y de Estudios
Avanzados del I.P.N.\\
Apdo. Post. 14-740, 07000, M\'exico, D.F., M\'exico\\
$^2$ Facultad de Ciencias F\'{\i}sico-Matem\'aticas \\
Benem\'erita Universidad  Aut\'onoma  de  Puebla,
72570,  Puebla,  Pue.,  M\'exico}

\begin{abstract}
The smallness of the $\theta_{13}$ mixing angle as observed in neutrino oscillation experiments
can be understood through an approximated  $\mu - \tau$ 
exchange symmetry in the neutrino mass matrix. Using recent oscillation 
neutrino 
data,
but assuming no \textit{CP} violation,  we
study  $\mu-\tau$ breaking parameter space to establish the conditions under 
which
such a breaking could have a perturbative origin. According to the so-obtained 
conditions, we 
suggest that a sterile neutrino, matching LSND/MiniBooNE neutrino oscillation 
results,
could provide the necessary ingredients to properly fix atmospheric and 
$\theta_{13}$
mixing angles to observable values, without exceeding the sterile neutrino 
fraction bound in
solar oscillations. In such a scenario, we analyze the general effect of a 
fourth neutrino on the prediction for the
effective $m_{ee}$ majorana mass parameter.

\end{abstract}

\maketitle
\section{Introduction}

Neutrino oscillation experiments, using solar, atmospheric, reactor, and 
accelerator neutrinos, provide
compelling evidence in favor of  nonzero
neutrino masses and  mixings \cite{nuglobal,PDG}. 
With the exception of LSND~\cite{LSND}, MiniBooNe~\cite{MB}, and 
a recent reanalysis of the flux in some short baseline experiments 
\cite{short-baseline}, all  
existing neutrino oscillation data can be described, and understood, assuming  
the mixing of only three flavor (standard) neutrinos.
Within this framework, data indicate that two of the three neutrino mass 
eigenstates, $\nu_{1,2}$, have a squared mass difference given by 
$\Delta m_{21}^2= m_2^2 -m_1^2=\Delta m_{sol}^2 \sim 7.5 \times 10^{-5} eV^2$, 
whereas the third one, $\nu_3$, is separated from the $\nu_1 -\nu_2$ pair by a 
splitting given by $|\Delta m_{31}^2| \sim \Delta m_{ATM}^2 \sim 2.5\times 
10^{-3} eV^2$. However, 
the sign in $\Delta m_{31}^2= m_3^2 -m_1^2$, and therefore the neutrino mass 
hierarchy 
pattern, is still unknown.

Unlike the quark sector where mixing angles are all small, the measured mixings 
in oscillation experiments are large, except for $\theta_{13}$, which has been 
found to be rather small. In the standard parametrization, mixings are 
given by the Pontecorvo–Maki–Nakagawa–Sakata (PMNS) matrix \cite{Pontecorvo-57-58, 
MNS},
\begin{equation}
\label{PMNSmatriz}
U_{PMNS} =  \left( \begin{array}{ccc} 
c_{12} c_{13} & s_{12}c_{13} & s_{13} e^{-i\delta_{CP}} \\
- s_{12} c_{23} - c_{12}s_{23}s_{13}e^{i\delta_{CP}}   & 
c_{12}c_{23} - s_{12}s_{23}s_{13}e^{i\delta_{CP}} & s_{23}c_{13} \\
s_{12}s_{23} - c_{12}c_{23}s_{13}e^{i\delta_{CP}} & 
- c_{12}s_{23} -c_{23}s_{12}s_{13}e^{i\delta_{CP}} & c_{23}c_{13} 
\end{array} \right)\cdot K ~, 
\end{equation}
where $c_{ij}$ and $s_{ij}$ stand for $\cos \theta_{ij}$ and  
$\sin \theta_{ij}$, respectively,  of  the mixing angles given as 
$\theta_{12}$, $\theta_{13}$, and $\theta_{23}$. Here, $\delta_{CP}$ is the
Dirac \textit{CP} phase, whereas
$K=\text{Diag}(e^{i\beta_1/2},e^{i\beta_2/2}, 1)$ is a diagonal matrix 
containing two Majorana phases which do not contribute to neutrino 
oscillations. Because of the clear hierarchy in oscillation mass scales, where 
$\Delta m_{ATM}^2>>\Delta m_{sol}^2$ and the smallness of 
$\theta_{13}$, it is possible to make a direct identification of above mixings 
with the ones used in a simple two neutrino oscillation analysis of the data. 
This feature will be useful later on for theoretical approximations. 
Global fits with all three neutrinos indicate that~\cite{nuglobal,PDG} 
$\sin^2\theta_{12} \equiv \sin^2\theta_\odot\approx 0.308\pm 0.017$, 
$\sin^2\theta_{23}\equiv \sin^2\theta_{ATM}\approx 0.437^{+0.033}_{-0.023}~ 
(0.455^{+0.039}_{-0.031})$, and $\sin^2\theta_{13}\approx 
0.0234^{+0.0020}_{-0.0019} ~ (0.0240^{+0.0019}_{-0.0022})$, for normal 
(inverted) hierarchy. $\delta_{CP}$, on the other hand, has not been 
determined well so far.

As in the quark sector, the matrix in Eq.~(\ref{PMNSmatriz}) actually 
encodes mixings that are independently used to diagonalize both charged 
and neutral lepton masses. Nevertheless, 
it is always possible to rotate any lepton basis into that where both
charged lepton masses and weak interactions are simultaneously diagonal. 
In such a basis flavor associated to  $e$, $\mu$, and
$\tau$, labels became transparent, and, furthermore, the PMNS
matrix becomes the one that diagonalizes neutrino masses, given in
general by the effective operator
\be
(M_\nu)_{\alpha\beta} \bar{\nu}_{\alpha L} (\nu_{\beta L})^c 
+ h.c., 
\ee
such that $U_{PMNS} = U_\nu\cdot K$. Therefore, the neutrino mass matrix can be
written in terms of diagonal (complex) masses, $M_{diag} = 
\text{Diag}\{ m_1 e^{i\beta_1},m_2 e^{i\beta_2}, m_3 \}$,  simply as
\be 
M_\nu = U_\nu\cdot M_{diag} \cdot U_\nu^T.
\label{mnu}
\ee
We will work in such a base hereafter. 
It is worth noticing that, while the observed $\theta_{13}$ is close to
zero, although non-null, $\theta_{ATM}$ 
is close to its maximal value, $\pi/4$. 
Certainly, neither of the central values of these angles is in such critical
values; however, it is intriguing to observe that, regardless of the hierarchy, 
it is possible to establish the approximated empirical relation
\be
1/2 - \sin^2\theta_{ATM}\approx \sin\theta_{13}/\text{few}~, 
\label{phen}
\ee
which suggests that the deviation of $\theta_{ATM}$ 
from its maximal value, $\Delta\theta=\pi/4 -|\theta_{ATM}|$,  could 
somehow be correlated to the nonzero value of $\theta_{13}$.
That would be the case if both parameters share the same physical origin. 
As a matter of fact, in the weak flavor basis we have chosen, it is easy to 
see that null values of  $\Delta\theta$ and $\theta_{13}$ do increase the 
symmetry in the mass neutrino sector, by exhibiting a discrete 
$\mu- \tau$ exchange symmetry~\cite{mutau}. As a consequence, observed values 
of these mixings could be understood as a result of the breaking of $\mu-\tau$ 
symmetry.
This fact has inspired many theoretical studies  in the last 
years~\cite{mutau,others,mtseesaw,gupta}, but little attention has been paid to 
exploring models that might provide a physical reason for such a breaking. 
That is the main question we shall address in the present paper by 
suggesting the mixing with a fourth sterile neutrino, that also accounts for 
LSND/MiniBooNE observed oscillations, as the natural source for the violation 
of 
$\mu-\tau$ symmetry. This idea has been explored in Refs. \cite{rodejohann,merle}, although our general scope in here is quite different.

The paper is arranged as follows. To clearly establish our sterile neutrino 
hypothesis, we start by revisiting  $\mu-\tau$ symmetry and parametrizing  
its 
breaking. Next, we use experimental results on neutrino masses and mixings 
to explore breaking parameter space, assuming \textit{CP} conservation for 
simplicity, to show that relatively small parameters, and therefore 
perturbative approximations, are well allowed by the data, provided standard 
neutrino masses are almost degenerate. As we will argue, the order of magnitude 
of such parameters suggests that the naive physical mass scale associated to 
$\mu-\tau$ breaking could straightforwardly be identified as the LSND/MiniBooNE 
scale. Hence, we elaborate a general model for neutrino masses and mixings, 
including a sterile neutrino, and explore  
the feasibility that the source of the breaking came from the sterile neutrino 
sector, the nonsymmetric couplings of which
provide for the necessary ingredients to fix all 
mixings in the model. As we will show, 
there is indeed  a non-null region in 
parameter space where all  experimental observables can be accommodated within 
one standard deviation.
Furthermore, we calculate the sterile fraction in solar neutrinos  
predicted by the model and discuss the impact of our sterile neutrino model in 
neutrinoless double beta decay experiments. Finally, we present our conclusions.

\section{neutrino mixings and $\mu -\tau$ symmetry}

First of all, let us remark that in the theoretical limit of
null $\theta_{13}$ and  
$\theta_{ATM}= -\pi/4$, with only three 
standard flavor neutrinos, there is not a Dirac 
\textit{CP} phase and mixing matrix $U_\nu$ becomes the bimaximal 
mixing form
\begin{equation}\label{Ubm}
U_{BM}=  \left( \begin{array}{ccc} 
\cos\varphi_{12}                     & \sin\varphi_{12}                  & 0 \\
\frac{- \sin\varphi_{12}}{\sqrt{2}}  & \frac{\cos\varphi_{12}}{\sqrt{2}} & 
\frac{-1}{\sqrt{2}} \\
\frac{- \sin\varphi_{12}}{\sqrt{2}}  & \frac{\cos\varphi_{12}}{\sqrt{2}} &  
\frac{1}{\sqrt{2}}
\end{array}\right),
\end{equation}
where the only undefined mixing corresponds to $\varphi_{12}$, which 
eventually, upon small corrections, will become the solar mixing.
Using this matrix within Eq.~(\ref{mnu}), one can read out the general form of 
the mass terms, which turn out to be symmetric 
under the exchange of $\mu$ and $\tau$ labels. Indeed,  by defining the 
mass matrix elements as $m^0_{\alpha\beta} = (M_\nu)_{\alpha\beta}$, one obtains 
\bea
m^0_{ee}&=& m_{1}\cos^{2}\varphi_{12}+m_{2}\sin^{2}\varphi_{12};\nonumber\\
m^0_{e\mu}=m^0_{e\tau}&=&\frac{\sin2\varphi_{12}}{\sqrt{8}}\left(m_{2}-m_{1}
\right);\nonumber\\
m^0_{\mu\tau}&=&\frac{1}{2}\left(m_{1}\sin^2\varphi_{12}+m_{2}\cos^2\varphi_{
12} -m_ { 3 } \right);\nonumber\\
m^0_{\mu\mu}=m^0_{\tau\tau}&=&
\frac{1}{2}\left(m_{1}\sin^{2}\varphi_{12}+m_{2}\cos^2 \varphi_{12} 
+m_{3}\right)~;\label{mterms}
\eea
where Majorana phases are to be understood.

Conversely, in the ``top-down'' approximation, the so-called $\mu - \tau$ 
symmetry~\cite{mutau} is expressed as the starting point on mass terms by 
two general conditions given as 
$m^0_{e\mu}=m^0_{e\tau}$ and $m^0_{\mu\mu}=m^0_{\tau\tau}$, which reduce the number 
of free mass parameters to $4$. Thus, in the limit of exact symmetry, one  
obtains the predictions for mass eigenvalues
\bea
m_{1}&=&m^0_{ee}-\sqrt{2}{m}^0_{e\mu}\tan\varphi_{12},\nonumber \\
m_{2}&=&m^0_{ee}+\sqrt{2}{m}^0_{e\mu}\cot\varphi_{12},\nonumber \\
m_{3}&=&{m}^0_{\mu\mu}-m^0_{\mu\tau}~.\label{masseigen}
\eea
where $1-2$ mixing is given by
\be
\tan2\varphi_{12}=
\sqrt{8}\left[\frac{{m}^0_{e\mu}}{{m}^0_{\mu\mu}+\left(m^0_{\mu\tau}-m^0_{ee}
\right)}\right]~.\label{12mix}
\ee
Besides,  null values for $\theta_{13}$ and $\Delta\theta$ are predicted.
However, as already mentioned, this last  is not the case from experimental 
results. Nevertheless, 
$\mu - \tau$ can still be assumed as a rather approximated symmetry in the neutrino 
sector, such that understanding the sources that contribute to its breaking may 
enlighten the origin of neutrino mixings. Next, we will elaborate on the
parametrization for the breaking of $\mu-\tau$ symmetry.

In general, any generic neutrino mass matrix can always be parametrized in 
terms of a symmetric part plus a correction that explicitly breaks the 
symmetry, by $M_\nu = M_{\mu - \tau} + \delta M,$ where $ M_{\mu-\tau}$ 
does posses a $\mu-\tau$ symmetry, whereas $\delta M$ is defined 
by only two nonzero elements, 
\begin{equation}
\delta M = \left( \begin{array}{ccc} 
0  & 0  & \delta  \\
0 & 0 & 0 \\
\delta & 0 & \epsilon
\end{array} \right)~, 
\end{equation}
where breaking parameters are, clearly, defined as $\delta = m_{e\tau} - 
m_{e\mu}$ and $\epsilon = m_{\tau \tau} - m_{\mu \mu}$. In this line of 
thought, understanding the origin of these parameters is a key to understanding 
$\theta_{13}$ and $\theta_{ATM}$.

Assuming that these are 
relatively small parameters, in comparison to $m_{e\mu}$ and $m_{\mu\mu}$ 
respectively,  observable mixing angles are estimated, in the absence of \textit{CP} 
violation, to satisfy
\bea
\tan2\theta_{\odot}&\approx&\sqrt{8}\left[\frac{\bar{m}_{e\mu}}{\bar{m}_{\mu\mu}
+\left(m_{\mu\tau}-m_{ee}\right)}\right];\label{mixings}\\
\sin\theta_{13}
&\approx&\frac{1}{\sqrt{8}}\left[\frac{2m_{\mu\tau}\delta-\epsilon\bar{m}_{e\mu}
}{\bar{m}_{e\mu}^{2}+m_{\mu\tau}\left(\bar{m}_{\mu\mu}-m_{\mu\tau}-m_{ee}\right)
} \right];\nonumber\\
\sin\Delta\theta
&\approx&\frac{1}{4}\left[\frac{\epsilon\left(m_{\mu\tau}+  
m_{ee}-\bar{m}_{\mu\mu}\right)+2\bar{m}_
{e\mu}\delta}{\bar{m}_{e\mu}^{2}+m_{\mu\tau}\left(\bar{m}
_{\mu\mu}-m_{\mu\tau}-m_{ee}\right)}\right]~,
\nonumber
\eea
where $\bar{m}_{e\mu}\equiv \left(m_{e\mu}+m_{e\tau}\right)/2$ and 
$\bar{m}_{\mu\mu}\equiv \left(m_{\mu\mu}+m_{\tau\tau}\right)/2$~. Notice 
that for small$\Delta\theta$ , one would have that 
$\sin\Delta\theta\approx 1/2 - \sin^2\theta_{ATM}$, which jointly to 
$\sin\theta_{13}$ would be given by linear relations in terms of $\epsilon$ and 
$\delta$.
Of course, the former expressions are first-order calculations that would 
provide a 
good approximation, provided the breaking parameters are small enough. It 
is  remarkable, though, that solar mixing turns out to have a similar expression 
to that obtained in the exact symmetric limit.

Since we already have quite more precise information about the mixing angles,  
it seems interesting to look at the parameters the other way around, 
by addressing 
the theoretical question regarding how good $\mu-\tau$ is as an approximated 
symmetry, that is, to obtain information about the relative size of the 
breaking parameters, as a way to search for hints of any possible physics 
lying 
beneath them. In particular, for instance, knowing to what extent 
$\delta M$ could be treated as a perturbation  
could give a hint toward knowing how far in the energy scale the breaking source 
lies away from the overall active neutrino mass scale. This possibility is  in 
itself an interesting one, and our main goal on the following 
discussion will be to explore under which conditions one could achieve a 
perturbative breaking of $\mu-\tau$, meaning acceptable small values for the 
breaking parameters $\epsilon$ and $\delta$.

Early work has shown that the source of such a breaking cannot 
come from within the Standard Model physics, where the only breaking source is 
the $\mu$ 
and $\tau$ mass difference~\cite{mtseesaw}. As a matter of fact, this 
charged lepton mass difference is indeed communicated through charged weak 
interactions to the neutrino 
sector, becoming, upon radiative corrections, a source for nonzero $\delta M$.
Nevertheless, such a correction turns out to be too small to account for 
observed mixings. Therefore, we are moved to assume that there should be a 
breaking sector out of the Standard Model. 

Without relying on any approximation, one could make a 
direct reconstruction of  the mass matrix in Eq.~(\ref{mnu}) and 
thus of the actual values for $\delta$ and $\epsilon$ parameters. To 
this aim, however, one would require knowledge of the mass spectrum, which we 
do 
not have so far. What we do have, instead, are the two values for mass squared 
differences involved in neutrino oscillations, $\Delta m^2_{sol}$ and $\Delta 
m^2_{ATM}$. Thus, one mass parameter in the spectrum, which we 
take as the lightest absolute neutrino mass, aside from the 
relative sign of mass eigenvalues, would remain as free 
parameters. Notice that by the last we mean to take  Majorana phases to 
be either $0$ or $\pi$ so that they provide just a relative sign for the 
masses. Dirac \textit{CP} phase we will assume hereafter to be zero. In these terms, we 
rewrite the absolute mass eigenvalues as
\begin{eqnarray}
\label{eigenvalues}
|m_2| = \sqrt{m_0^2 +  \Delta m_{sol}^2} ~&\text{~and}& ~
|m_3|= \sqrt{m_0^2 + |\Delta m_{ATM}^2|}~~ \text{for NH}. \\
|m_1| = \sqrt{m_0^2 + |\Delta m_{ATM}^2|} ~&\text{~and}& ~
|m_2|= \sqrt{m_0^2 +| \Delta m_{ATM}^2|+ \Delta m_{sol}^2} ~~
\text{ for IH}. \nonumber
\end{eqnarray}
Note that, in above, the lightest mass eigenstate,  $m_0$, becomes 
$m_1$ for the normal mass hierarchy (NH) and $m_3$ for the inverted 
mass hierarchy (IH). 
Next, to proceed with our analysis, we define, without approximations, the 
dimensionless parameters
\begin{eqnarray}
\label{deltaepsilon2}
 \hat \delta \equiv \frac{\delta}{m_{e \mu}} &= 
\frac{ \sum_i (U_{ei} U_{\tau i}-  U_{ei} U_{\mu i} ) m_i}{\sum_i U_{ei} U_{\mu 
i} m_i }
 \nonumber \\
\hat \epsilon \equiv \frac{\epsilon}{m_{\mu \mu}} &= 
\frac{\sum_i (U_{\tau i} U_{\tau i}-  
U_{\mu i} U_{\mu i}) m_i}{\sum_i U_{\mu i} U_{\mu i} m_i}~,
\end{eqnarray}
where the right-hand-sides have been written according to Eq.~(\ref{mnu}). Combined with 
Eq.~(\ref{eigenvalues}), the last expressions give the dimensionless parameters in 
terms of observed mixing angles, oscillation mass scales, and the absolute 
scale of neutrino masses, $m_0$, as the only free parameter.  
Next, let us perform an approximated analytical analysis of 
the expressions in Eq.~(\ref{deltaepsilon2}) by considering all four independent 
combinations of mass signs: (i) $m_{1,2,3}>0$, (ii) $m_{1,2}<0$ but $m_3>0$, 
(iii) $m_{1,3}>0$ but $m_2<0$, and (iv) $m_1<0$ but $m_{2,3}>0$. Those can be 
written in the suitable form 
\begin{eqnarray}\label{dehat}
\hat{\delta} &=& \frac{y_{-} f  s_{13}+y_{+}}{1+f \ s_{13} \tan~\theta_{23}}~,  
\nonumber \\
\hat{\epsilon} &=& \frac{g \cos~2\theta_{23} - s_{13}h}{1+g s_{23}^2+s_{13}h/2} ~,
\end{eqnarray}
where
\begin{eqnarray}
y_{\pm} &=& \frac{c_{23}\pm s_{23}}{c_{23}}, \nonumber \\
f &=& \frac{m_1 c_{12}^2 + \sigma m_2  s_{12}^2 -\Sigma m_3 }
           {c_{12}s_{12}(m_1 -\sigma m_2 )}~, \nonumber \\
g &=&\frac{m_1( c_{12}^2 s_{13}^2 - s_{12}^2)+\sigma m_2( s_{12}^2 s_{13}^2 
-c_{12}^2) +\Sigma m_3 c_{13}^2 }{m_1 s_{12}^2 + \sigma m_2 c_{12}^2}~, 
\nonumber\\
h &=&\frac{(m_1 +\sigma m_2 )\sin~2\theta_{23}~ 
\sin~2\theta_{12}}{m_1  s_{12}^2 +\sigma m_2  c_{12}^2 } ~,
\end{eqnarray}
with the conventions $\sigma=+$, $\Sigma=\pm$ for  cases i and ii 
and $\sigma=-$, $\Sigma=\pm$ for  cases iii and iv, respectively. 
As one can see from these expressions,  $\hat \delta$ and $\hat \epsilon$ in 
Eq.~(\ref{dehat}) become zero when $\theta_{13}=0$ and $\theta_{23}=-\pi/4$, as 
expected from exact $\mu-\tau$ symmetry. In the following, let  us 
first examine under which considerations $\hat \delta \ll 1$, and latter 
on, we will analyze the behavior of $\hat \epsilon$ in such cases. In the 
three approaches given by the hierarchies, and using the central values for the 
current mixing parameters, we have
\begin{itemize}
\item  For NH, $m_1 \ll m_2 \approx \sqrt{\Delta m_{sol}^2} \ll m_3 
\approx \sqrt{\Delta m_{ATM}^2}$ , and thus
\begin{equation}
f \approx \frac{\Sigma}{\sigma s_{12}~c_{12}} 
\sqrt{\frac{\Delta m_{ATM}^2}{\Delta m_{sol}^2}} \left(1-\frac{\sigma 
s_{12}^2}{\Sigma} \sqrt{\frac{\Delta m_{sol}^2}{\Delta m_{ATM}^2}} \right)~, 
~|f|\sim 12.5 ~,
\end{equation}
which implies $|\hat \delta| \sim 3.26$, discarding NH for any mass sign 
combinations. 
\item For IH, $m_1\approx \sqrt{\Delta m_{ATM}^2}~,~  m_2 \approx 
\sqrt{\Delta m_{sol}^2+ \Delta m_{ATM}^2} \gg m_3 $ , which gives
\begin{equation}
f \approx \frac{c_{12}^2 + \sigma s_{12}^2 + 
\frac{\sigma s_{12}^2}{2}\frac{\Delta m_{sol}^2}{\Delta 
m_{ATM}^2}}{s_{12}~c_{12}\left(1-\sigma - \frac{\sigma}{2} \frac{\Delta 
m^2_{sol}}{\Delta m^2_{ATM}} \right)} ~. 
\end{equation}
For cases i and ii, we have $|f|\sim 10^{2}$, 
such that $|\hat \delta| \sim 2$, whereas in cases iii and iv, we obtain 
$|f|\sim 1$ and hence $|\hat \delta| \sim 0.1$. Therefore, cases i 
and ii again seem to be ruled out.
\item Finally, for degenerated hierarchy (DH), we get\\
\begin{equation}
f\approx \frac{c_{12}^2+\sigma s_{12}^2-\Sigma - 
\frac{\Sigma}{2}\frac{\Delta m^2_{ATM}}{m_0^2}}{c_{12}^2 s_{12}^2 
\left(1-\sigma - \frac{\sigma}{2}\frac{\Delta m^2_{sol}}{m_0^2}  \right)} ~.
\end{equation}
Cases i and ii give $|f| \gtrsim 10^2$, which now implies $|\hat \delta| 
\gtrsim 15$, while for cases iii and iv, one gets $|f| \lesssim 1 $  and 
$|\hat \delta| \lesssim 0.1$. Therefore, in DH, cases i and ii are once 
more disfavored. 
\end{itemize}
The approximations taken above suggest 
that only cases iii and iv, in the IH and DH, allow for small values of 
$\hat \delta$. In the IH, a similar analysis, after some 
algebra, gives $\hat \epsilon \sim 1$, for the iii and iv combinations, 
whereas in the DH, we obtained $\hat \epsilon \sim 1$ for case iii and $\hat 
\epsilon \sim 0.4$ for case iv. 
Thus, our analysis indicates that the only fairly perturbative 
case would occur in the DH for the signs combination that corresponds to   
$m_1<0$ and $m_{2,3}>0$.


\begin{figure}
\begin{small}
\begin{center} 
\includegraphics[scale=0.5]{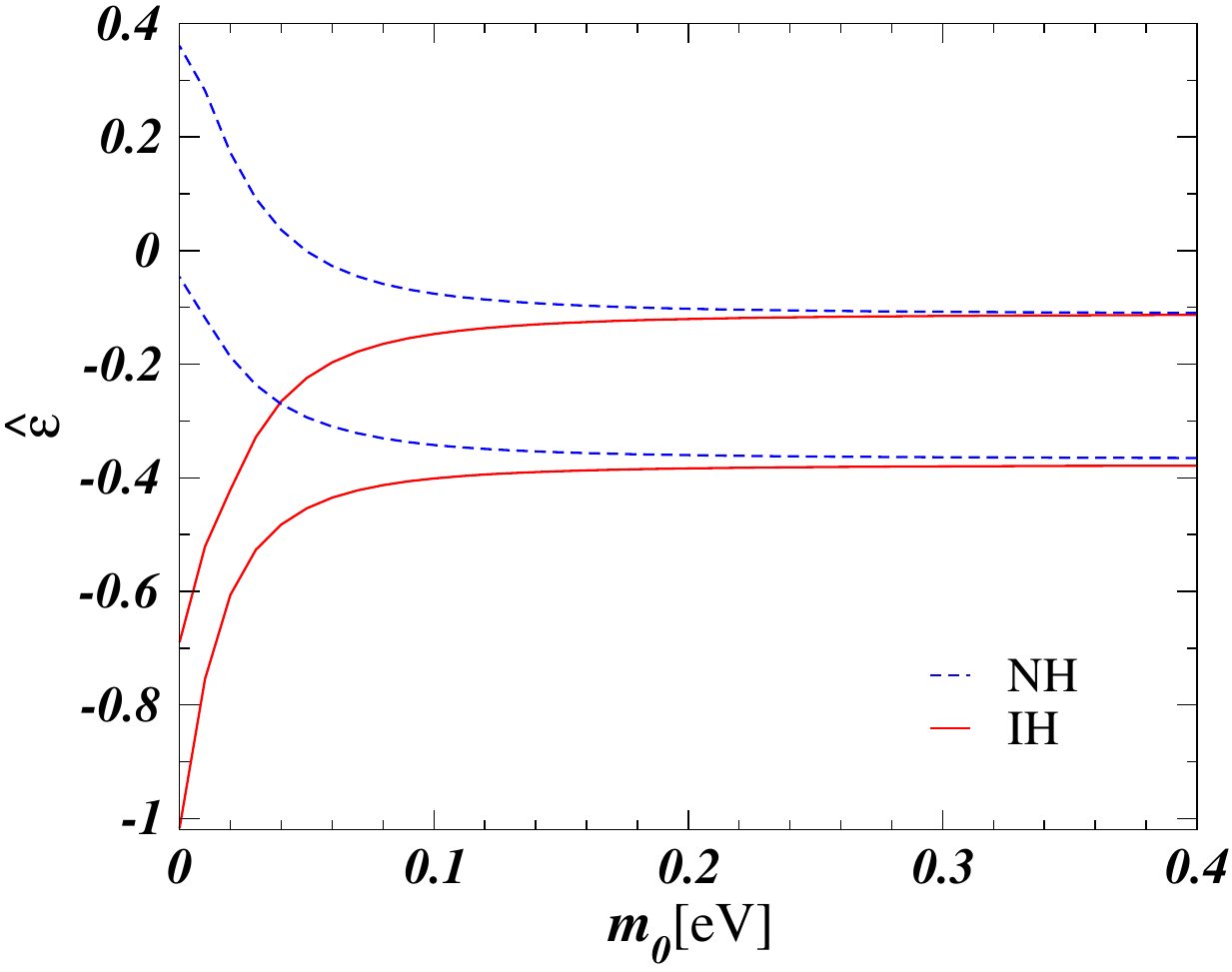} ~
\includegraphics[scale=0.5]{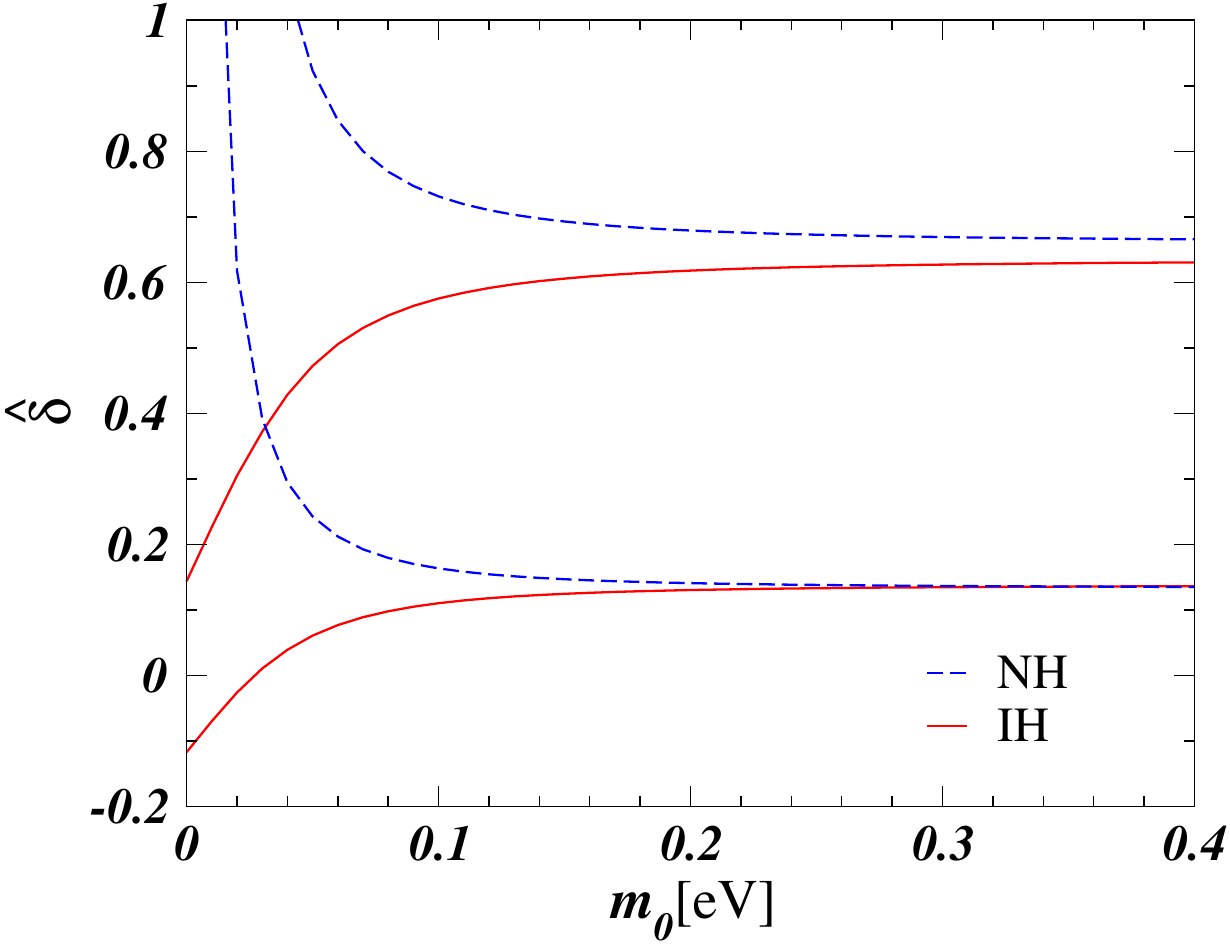}
\caption{ \label{epsdelnorinv}
\small{One sigma regions for the allowed values of dimensionless $\mu-\tau$ 
breaking parameters, $\hat\epsilon$ and $\hat \delta$,  
as a function of the lightest neutrino mass, $m_0$, for inverted 
(red line) and  normal (blue dashed line) hierarchy, in the case where 
$m_1 < 0$, $m_{2,3} > 0$.}} 
\end{center}
\end{small}
\end{figure} 

After a complete numerical analysis of the parameter space 
allowed by 
data (without any approximation), in all four mass sign independent combinations, 
it was found that, while in all possible cases there is always 
a solution with nonzero values for either of, or simultaneously both, 
the $\hat\delta$ and $\hat \epsilon$ parameters, 
the only case one might consider as fairly perturbative corresponds to 
$m_1 < 0$ and $m_{2,3} > 0$ (regardless of the hierarchy). 
This is consistent with our previous analysis. 
The allowed one sigma region for both breaking parameters in this case is 
depicted in Fig.~\ref{epsdelnorinv}. Moreover, as one can see from this figure, 
only for almost degenerate neutrinos, where $m_0 > 0.1$~eV, is it possible to 
actually pick up relatively small values for $\hat\delta$ and $\hat \epsilon$ 
to comply with the expectation of a perturbative origin. Interestingly enough, 
none of the breaking parameters is null within such an allowed region. We must 
mention that our results are consistent with those obtained in the general 
analysis made in Ref.~\cite{gupta}, although our general scope here is quite 
different.

Naively, if one takes, for instance, $\hat\epsilon \simeq \hat\delta \simeq 0.2$ 
or so, valid in the whole region for $m_0$ above $0.1~eV$ for $m_0$,  the 
approximation 
we used to derive the mixings in Eq.~(\ref{mixings}) would be quite well 
justified,  and so would be, to the numerical extent,  the 
geometrical relation among mixings given in Eq.~(\ref{phen}).

\section{ $\mu -\tau $  breaking from a sterile neutrino}

There are at least two possible approximations one can make to 
explore the physics beyond standard model that is responsible of generating 
the breaking of $\mu-\tau$ symmetry. Either this lies close to the same 
physics that is responsible for the smallness of neutrino masses, in which case 
one has to probably go for model building to explore concrete possibilities, or
it is the consequence of the mixing with a sector that does not comply 
with the symmetry. An example of the latter is the mixing corrections induced 
through 
radiative processes and  due to the explicit violation of the symmetry in the 
charged lepton masses. As mentioned already, this is too small to account 
for 
the observed effect in neutrino mixings.
Another quite straightforward candidate for this would be a sterile 
neutrino, which by definition does not have weak interactions, and thus it has 
no (e, mu, or tau) lepton flavor.
This last possibility is much more intriguing, because the 
not-so-small parameters that are required to understand the mixings do suggest 
that such a new sector cannot be too far away from the standard neutrino mass scale.
Actually, by assuming that the breaking parameters are somehow generated at a 
given larger scale, $m_s$, and naively  
taking the perturbations that break the symmetry as given 
in terms of the ratio among the involved scales, which means that 
$\hat\epsilon~,\hat\delta \approx m_\nu/m_s$,  then the mass scale of the 
sterile neutrino should be just about the 
$eV$ scale, precisely as suggested by LSND/MiniBooNe results. Next, we will 
analyze in detail such a possibility.

To be specific in our analysis, we assume a single light sterile neutrino 
and 
consider its most general mass terms, including the mixing with the standard 
active neutrino sector. Notice, however, that we shall be working in a  
$3+1$ neutrino mixing scheme in which 
the fourth  neutrino (predominantly sterile) is isolated from the block of 
three active  flavor neutrinos by the mass gap $\Delta m^2_{LSND}\approx 
(0.4-10)~\text{eV}^2$.
Hence, in 
the basis 
($\nu_e,~\nu_\mu,~\nu_\tau,~\nu_s$), the mass matrix can be written as
\begin{equation}
\label{matrizdemasa4x4compactaaprox}
{\cal M} = \left( \begin{array}{cc} 
M_{\mu - \tau}   &  \overrightarrow{\alpha} m_{s} \\
\overrightarrow{\alpha}^T m_{s} &   m_{s} 
\end{array} \right),
\end{equation}
where  $m_s$ is the Majorana mass for the sterile neutrino and the vector
$\overrightarrow{\alpha}^T= (\alpha_e, \alpha_\mu, \alpha_\tau)$
denotes the active-sterile mixing masses in units of $m_s$. 
Specific structures of this vector could have consequences for model builders, 
as discussed in Ref. \cite{merle}. Next, let us assume
that $\alpha_\ell \ll 1$  and $m_s \gg m_{\alpha\beta}$, where $m_{\alpha\beta}$
are the elements of the active and explicitly symmetric flavor matrix 
$M_{\mu - \tau}$~. 
Clearly, if $\alpha_\mu = \alpha_\tau$, the whole sector 
would be invariant under $\mu-\tau$ symmetry, with the known consequences of it 
for active neutrino mixings. We will not assume so, and thus the model will 
have a single effective parameter for the breaking of the symmetry given by the 
coupling differences $\Delta\alpha= \alpha_\tau-\alpha_\mu$. Nevertheless, one 
would find it useful to keep track of the independent $\alpha$'s  along the 
calculations.

After decoupling $\nu_s$,  we get, at the lower-order approximation, an 
effective active flavor matrix, $M_\nu$,  the elements of which are given by the (low-energy) seesaw formula,
\begin{equation} \label{see-saw4} 
(M_\nu)_{\rho \delta} \simeq (M_{\mu 
- \tau})_{\rho \delta}-{\alpha}_{\rho} m_{s} 
{\alpha}_{\delta}^T~. 
\end{equation} 
It is clear that $M_\nu$ does not possess in general $\mu - \tau$ 
symmetry  due to the presence of the term $ {\alpha}_{\rho} m_{s} 
{\alpha}_{\delta}^T$.  It is important to notice that the last can always be 
separated in a symmetric plus a nonsymmetric part under the 
exchange of $\mu$ and $\tau$ indexes. 
The symmetric part, however, will only account for corrections, of second order 
in $\alpha_\ell$, to the mass spectrum and the solar mixing angle defined by 
$M_{\mu-\tau}$ alone. On the other hand, the nonsymmetric part would be the 
source for the breaking parameters defined in the previous section. It is 
actually easy to see that, without further approximations, one gets $\delta = 
\alpha_e\Delta\alpha m_s$  and 
$\epsilon= 2\bar\alpha_\mu\Delta\alpha m_s$, where, as before, 
$\bar\alpha_\mu= (\alpha_\mu+\alpha_\tau)/2$. It is  straightforward to show 
that, 
regardless of
hierarchy,  our now effective 
dimensionless breaking parameters are second order 
in $\alpha_\ell$, and for quasidegenerate neutrinos, they
can be approximated as
\bea
\label{delepsenterminosdealfa}
\hat \delta &\approx & 
\frac{\sqrt{2} m_s\alpha_e \Delta\alpha}
{m_0\sin 2 \varphi_{12}} ~,
\nonumber \\
\hat{\epsilon }&\approx & 
\frac{2 m_s \bar\alpha_\mu\Delta\alpha}{m_0\cos^2\varphi_{12}}  ~,
\eea
where the mixing $\varphi_{12}$ is the one involved in the diagonalization of 
the symmetric sector. 
Within a rough approximation, at lower order, one would have
$\varphi_{12}\sim\theta_{\odot}$. Therefore, to get an idea of the order of 
magnitude of the 
sterile to active neutrino couplings, one may take, for instance, $m_s\sim 1~eV$ 
$m_o\sim 0.2~eV$, and $\hat\epsilon\sim\hat\delta\sim 0.2$ which are 
consistent with the analysis in previous sections, to show that a solution the 
above formulas is found for $\alpha_e\sim 0.23$, $\bar\alpha_\mu\sim 0.16$, and 
$\Delta\alpha\sim 0.11$. Notice, however, that the effect of $\alpha_e$, even 
for $\Delta\alpha=0$, is to incorporate corrections to the mixing in the $1-2$ 
sector, and thus, a more accurate calculation is likely to modify these naive 
estimates.

Notice that, in getting the above results, it seems that 
three $\alpha_\ell$ couplings do contribute to only two effective breaking 
parameters, $\hat\delta$ and $\hat\epsilon$. Nevertheless, we would expect that 
any physical solution should at least be around above roughly estimated values 
for $\alpha_{\mu,\tau}$.  
To address this issue in a more reliable way, one should explicitly confront 
mass scales and mixing angles as obtained by the diagonalization of the 
complete  $3+1$ neutrino sector against measured experimental parameters. 
To this 
aim, let us first point out that the mass 
matrix given in Eq.~(\ref{matrizdemasa4x4compactaaprox})  contains eight 
independent parameters ($m_{\ell\ell'},~ \alpha_\ell$, and $m_s$), whereas we 
have knowledge of seven experimentally determined observables,  
enumerated as follows. From weak flavor 
oscillations one gets two squared mass 
scales, $\Delta m^2_{ATM}$ and $\Delta m^2_{sol}$, and three mixing angles,  
$\theta_{\odot}$, $\theta_{ATM}$, and $\theta_{13}$. Additionally, from 
LSND/MiniBooNe results, one gets two 
parameters, taken as a squared mass scale $\Delta m^2_{LSND}\approx \Delta 
m^2_{s\ell} \approx m_s$ and a mixing, $\theta_{e\mu}$. Therefore, there would 
be only one free parameter in the analysis, which we take 
as the lightest neutrino mass scale, $m_0$.
As already discussed, the consistency 
of our model with a perturbative treatment of the breaking of  $\mu-\tau$ 
symmetry requires $m_0$ to be within $0.1$ to $0.4~eV$, which corresponds  to degenerated hierarchy. 
This short range for 
$m_0$ will end up narrowing the allowed parameter space, as we will show below.

Next, for our analysis, we will take the intermediate neutrino mass eigenvalues 
as given in terms of the atmospheric and solar
scales by Eq.~(\ref{eigenvalues}). Moreover, following 
the outcome of the previous discussion, and considering  a perturbative
diagonalization of $\cal M$, one 
can see that active mass eigenvalues are well approximated (at lower 
order) by the eigenvalues of $M_{\mu-\tau}$, given in Eq.~(\ref{masseigen}), 
whereas $m_4\approx m_s$.
This leave us only with the question of constructing a self-consistent
system of equations to fit all experimental mixing angles with the remaining
parameters of the model. By considering the relevant effective oscillations in  
solar, atmospheric, and short baseline
experiments, one gets the general formulas
\begin{eqnarray} 
\label{constr1}
\sin^2 2\theta_\odot &=& 4 |\mathcal{U}_{e1}|^2 |\mathcal{U}_{e2}|^2 ~, \\
\label{constr2}
\sin^2 2\theta_{13} &=& 
4 \vert \mathcal{U}_{e3} \vert^2 ( \vert  \mathcal{U}_{e1} \vert^2 + 
\vert  \mathcal{U}_{e2} \vert^2)~, \\
\label{constr3}
\sin^2 2\theta_{ATM} &=& 
4 \vert \mathcal{U}_{\mu 3} \vert^2 \vert \mathcal{U}_{\tau 3}\vert^2 ~, \\
\label{constr4}
\sin^2 2\theta_{e\mu} &\simeq & 4 \vert \mathcal{U}_{e4} \vert^2 \vert
\mathcal{U}_{\mu 4} \vert^2~,
\end{eqnarray}
where $\mathcal{U}_{\alpha i}$,  for $i=1,2,3,4$, and $\alpha= e,
\mu,\tau, s$, stands for the elements of the general mixing matrix which 
diagonalizes $\cal M$. As it is well known, since we are neglecting \textit{CP} 
violation, the columns of $\cal U$ are given by the properly normalized 
eigenvectors and $\cal M$.
Here, we emphasize that the left-hand (lhs) sides of Eqs. (\ref{constr1}-\ref{constr3}) are 
known from usual neutrino oscillation experimental data, whereas the lhs of 
Eq.~(\ref{constr4}) comes from considering the allowed regions
of LSND and Mini-BooNE neutrino data \cite{3+1}, which we take as
\begin{equation}
\label{4ec}
\sin^2 2\theta_{e\mu}  = 0.0023, ~~~ \vert \Delta m_{41}^2 \vert = 0.89 eV^2.
\end{equation}

On the other hand, the entries in rhs of Eqs. (\ref{constr1}-\ref{constr4}), 
are given up to ${\cal O}(\alpha^2)$ by the following expressions:
\begin{eqnarray}
\label{Ue2}
{\cal U}_{e1}&\approx& c_{12}  +  
  s_{12} \frac{m_4^2}{m_{12}~m_{14}}\alpha_+\alpha_- - 
  \frac{c_{12}}{2}\frac{m_4^2}{(m_{14})^2}\alpha_-^2~, \label{Ue1}\\[1em]
{\cal U}_{e2} &\approx& s_{12} + c_{12} \frac{m_4^2}{m_{21}~m_{24}} 
  \alpha_+\alpha_- -
  \frac{s_{12}}{2}\frac{m_4^2}{(m_{24})^2}\alpha_+^2~, \\[1em]
\label{Ue3}
{\cal U}_{e3} &\approx& \frac{\Delta\alpha}{\sqrt{2}} \frac{m_4^2}{m_{34}}
  \left[  \frac{c_{12}}{m_{31}} \alpha_-  + \frac{s_{12}}{m_{32}}\alpha_+ 
  \right] ~,\\[1em]
\label{Ue4}
{\cal U}_{e4} &\approx& \frac{m_4}{m_{41}}\alpha_e ~,\\[1em]
\label{Uu3}
{\cal U}_{\mu 3} &\approx & -\frac{1}{\sqrt{2}} +
  \frac{\Delta\alpha}{2}\frac{m_4^2}{m_{34}} 
  \left[\frac{c_{12}}{m_{32}}\alpha_+ - \frac{s_{12}}{m_{31}} \alpha_-\right] 
  + \frac{(\Delta\alpha)^2}{4\sqrt{2}} \frac{m_4^2}{(m_{34})^2}~,\\[1em]
\label{Uu4} 
 {\cal U}_{\mu 4} &\approx& \frac{m_4}{m_{41}}\alpha_{\mu}~,\\[1em]
\label{Ut3}
{\cal U}_{\tau 3} &\approx& \frac{1}{\sqrt{2}} +
  \frac{\Delta\alpha}{2}\frac{m_4^2}{m_{34}} 
  \left[ \frac{c_{12}}{m_{32}}\alpha_+ -\frac{s_{12}}{m_{31}} \alpha_-\right] 
  - \frac{(\Delta\alpha)^2}{4\sqrt{2}} \frac{m_4^2}{(m_{34})^2}~.
\end{eqnarray}
Here, to simplify,  we have introduced the shorthand notation 
$m_{ij}=m_i-m_j$, for~ $i,j=1...4$, $\alpha_+=\alpha_e 
s_{12}+\sqrt{2} c_{12}\bar\alpha_\mu$ and  
$\alpha_-=\alpha_e c_{12}-\sqrt{2} s_{12}\bar\alpha_\mu$, where, as before, 
$c_{12}$ ($s_{12}$) stands for the cosine (sine) function of the free 
parametric angle, $\varphi_{12}$, defined by Eq.~(\ref{12mix}). 

As it is easy to see, one can use LSND/MiniBooNe 
mixing in order to solve for $\alpha_e$ in terms of $\alpha_\mu$, using 
Eq.~(\ref{constr4}). In the 
quasidegenerate neutrino scenario with sterile mass dominance that we are 
considering, this implies that 
 $ \sin2\theta_{e\mu}\approx 2|\alpha_\mu \alpha_e|$.
Numerically, this means that $|\alpha_\mu \alpha_e|\approx 0.02~$.
Similarly, in the same approximation, we obtain for the solar mixing
\beas
\sin^22\theta_\odot\approx \sin^22\varphi_{12}\cdot
\left[1 - (\alpha_e^2+2\bar\alpha_\mu^2) +
\frac{m_s}{m_o}\left(\cos2\varphi_{12}(\alpha_e^2 -2\bar\alpha_\mu^2) + 
\sqrt{8} \alpha_e\bar\alpha_\mu \cot2\varphi_{12}\right)\right]~,
\eeas
regardless of the hierarchy. It is worth noticing that the last expression does
depend on four effective parameters, $m_0$, $m_4$, and $\alpha_{\mu,\tau}$. 
In practice, since we are choosing $m_0$ in a given interval, this relation can 
be used to formally fix $\varphi_{12}$ mixing from the 
equations system, 
leaving us with only two relevant independent parameters: $\alpha_\mu$ and 
$\alpha_\tau$ couplings. Finally, these last parameters can be estimated (at least 
formally) from the formulas that give $\theta_{13}$ and atmospheric mixings, in 
Eqs.~(\ref{constr2}) and (\ref{constr3}), which at the lower 
order in the $\alpha$ parameters are written as
\bea
\sin^2 2\theta_{13} &\approx& 
\left(\frac{4\Delta\alpha m_s m_0}{\Delta m^2_{ATM}}\right)^2 
\left[\alpha_e \left(\frac{\Delta m^2_{ATM}}{4m_0^2}\pm\sin^2\varphi_{12}^2\right) 
\pm \frac{\bar\alpha_\mu\sin2\varphi_{12}}{\sqrt{2}}\right]^2~, 
\label{sin2t13}\\
\sin^22\theta_{ATM}&\approx& 1 - (\Delta\alpha)^2~.
\eea
The sign difference in Eq.~(\ref{sin2t13}) stands for normal and inverted 
hierarchies, respectively. Notice that $\sin^22\theta_{13}$ comes from 
corrections at the fourth order in $\alpha's$, although second order in $\Delta 
\alpha$, whereas atmospheric mixing gets a second-order correction, as suggested 
by the naive numerical expression in Eq.~(\ref{phen}).

\begin{figure}[t]
\begin{center}
\includegraphics[scale=0.5]{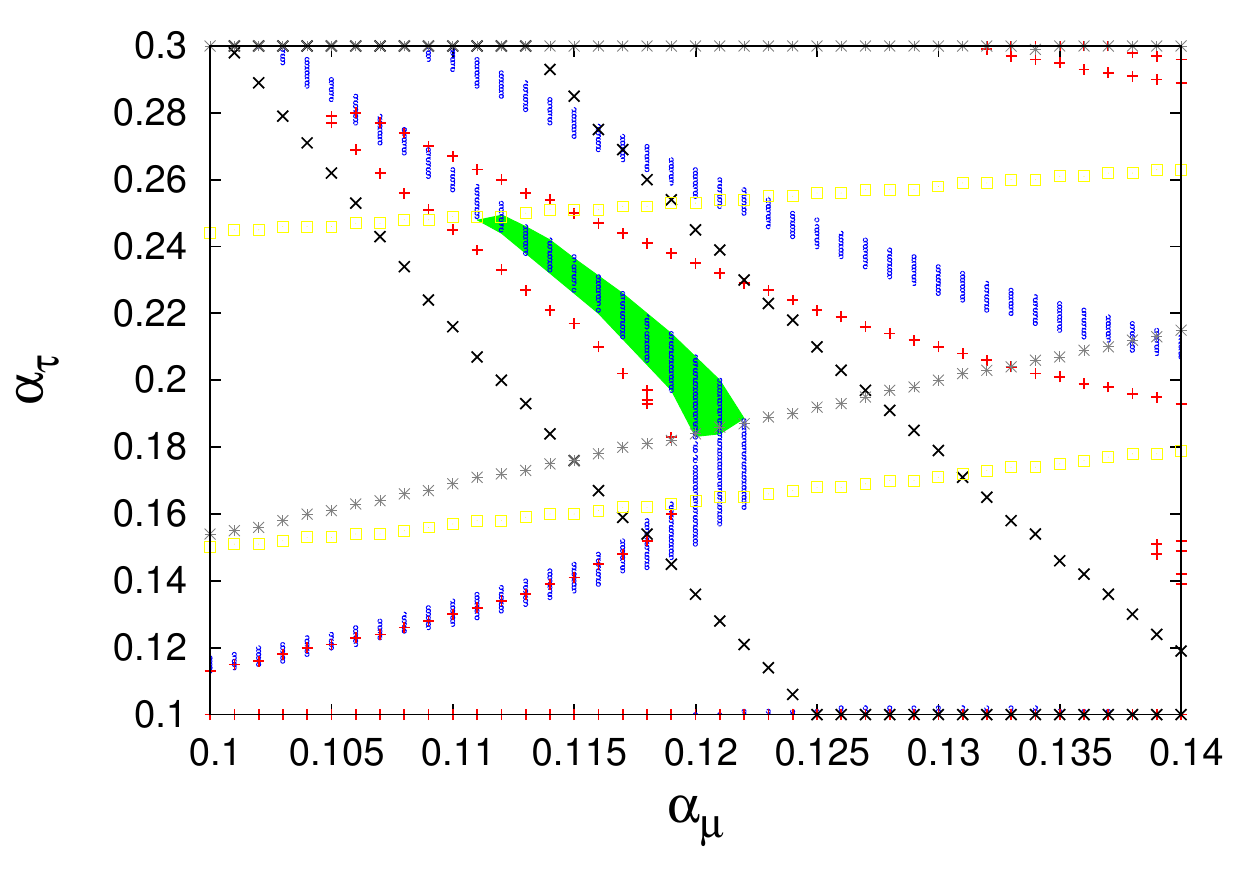} ~
\includegraphics[scale=0.5]{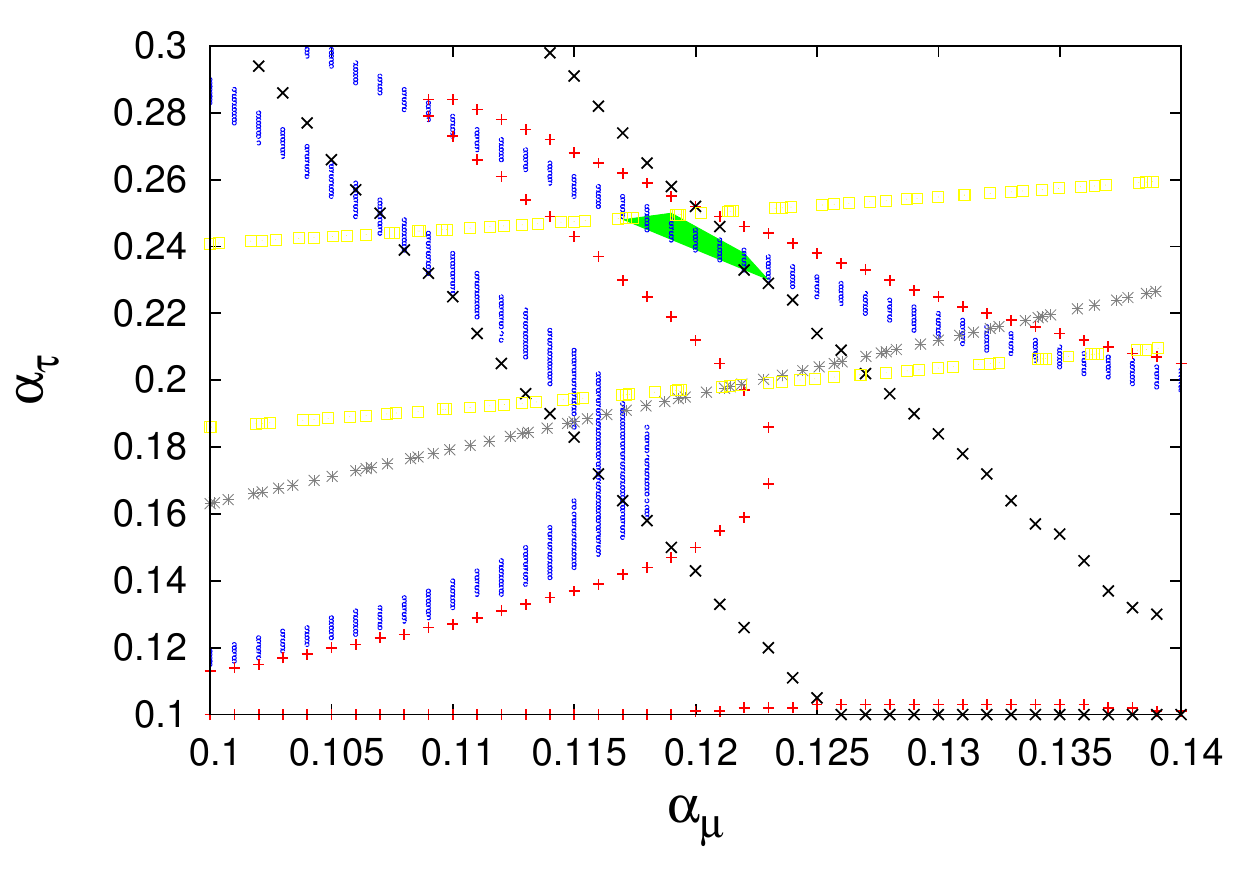}
\end{center}
\caption{ \label{regionsolution}
\small{
(Color on line) Allowed parameter space for $\alpha_\mu$ and $\alpha_\tau$ that 
is consistent 
with all oscillation neutrino data (shaded region), in the 3+1 scenario, 
within one sigma deviation for normal (lhs) and inverted (rhs) hierarchy, 
respectively. The doted (blue) region corresponds to the bounds given by 
$\sin^22\theta_{13}$, whereas  (red) crossed 
curves delimit the region for acceptable results of $\sin^22\theta_{ATM}$. 
The one sigma  $\sin^2\theta_{\odot}$ region is bounded by  (black) ex-marked 
curves. (Yellow) squared curves and (gray) star lines bound the
regions where $\hat \epsilon$ and $\hat\delta$ are small enough for the model 
to be perturbative.
} }
\end{figure}

Once we have some  understanding of the parameter correlations in the 
determination of the 
four observable mixings given in Eqs.~(\ref{constr1}) to (\ref{constr4}), we 
can now proceed with a numerical analysis of such a set of equations without 
further 
approximations, in order to explore and identify the allowed parameter space 
for 
$\alpha_\mu$ and $\alpha_\tau$ that gives consistent results for current 
experimental oscillation neutrino data, within one sigma 
deviations. Our results, for $m_0 = 0.2~eV$ and both the hierarchies, are 
presented in 
Fig.~\ref{regionsolution}, where we have scanned for appropriated values of 
$\alpha_\mu$ and $\alpha_\tau$ parameters for each mixing as independent, such 
that the consistent values are found in the overlapping of all regions (shaded 
area in the given plots) that give one sigma values for each standard mixing 
angle. 
In these same plots, we have also constrained the regions such that  $-0.4 
\lesssim \hat \epsilon \lesssim 0.3$ and $0.1 \lesssim \hat 
\delta \lesssim 0.6 $, to insure that the whole allowed parameter 
space be consistent with perturbative approximations.
The actual effect of including this last condition is to bound the allowed 
parameter space from the bottom and the top, as it can be seen on the given 
plots.

By picking up some allowed values for the $\alpha$ parameters, it is easy to 
reconstruct the four-by-four mass matrix to give an explicit numerical 
example for it. For a typical mass matrix obtained by this procedure, we 
consider   
\begin{equation}\label{M}
{\cal M}= \left(
\begin{array}{cccc}
 0.0247 & 0.0110 & 0.0110 & -0.1881 \\
 0.0110 & 0.0489 & -0.0205 & 0.1315 \\
 0.0110 & -0.0205 & 0.0489 & 0.0343 \\
 -0.1881 & 0.1315 & 0.0343 & 0.9428 \\
\end{array}
\right)~,
\end{equation}
for $\alpha_{e}=-0.1995$,~ $\alpha_{\mu}=0.1395$~, and $\alpha_{\tau}=0.0364$~.
This matrix leads to the exact active neutrino mixings, 
$\sin^2\theta_{\odot}=0.280$~, 
$\sin^2 \theta_{ATM}=0.379$, and $\sin^2 \theta_{13}=0.021$,
which are in good agreement with neutrino 
oscillation measured parameters, within two, three, and two sigma deviations, 
respectively.
Furthermore, we get for the LSND/MiniBooNe mixing $\sin^2 
\theta_{e\mu}=0.002$, in agreement with the fits of 
short-baseline neutrino oscillation data \cite{3+1}.
On the other hand, the corresponding squared mass differences obtained out of 
this example are $\Delta m_{sol}^2 = 7 \times 10^{-5}~eV^2$ and  
$\Delta m_{ATM}^2 = 2.6 \times 10^{-3}~eV^2$, whereas we get for the sterile 
mass eigenvalue $m_{s}=0.997~eV$, which are also consistent with observations. 
More accurate results could be obtained if higher-order 
corrections in $\alpha$ parameters are incorporated in the reconstruction of 
the mass matrix. Nevertheless, this numerical matrix serves as a good example 
to illustrate the mechanism we are exploring.


\section{sterile impact on other neutrino observables}

SK and SNO experiments have measured solar electron neutrino 
flux, $\Phi_{\nu_e}$, whereas SNO has also measured the total solar flux of 
active 
neutrinos, $\Phi_{\nu_e,\nu_\mu,\nu_\tau}$, using neutral current 
interactions.  These measurements are in good agreement with  the total 
neutrino flux, $\Phi_B$,  predicted by the solar model (see Ref. ~\cite{PDG} for 
further references), which 
can be used to constrain solar neutrino conversion into sterile 
neutrinos. Thus,  
assuming that solar neutrinos 
oscillate as $\nu_e \rightarrow \sin \alpha \nu_s + \cos \alpha 
\nu_{\mu,\tau}$, the 
sterile 
fraction $\eta_s \equiv \sin^2 \alpha$ is estimated to be \cite{sterilefraction}
\begin{equation}\label{boundsterile}
\eta_s \approx \frac{\Phi_B-\Phi_{\nu_e,\nu_\mu,\nu_\tau}}{\Phi_B-\Phi_{\nu_e}}
\approx 0 \pm 0.2~.
\end{equation}
Theoretically,  $\eta_s$ can be estimated in our model by using 
$\eta_s=\frac{P_{\nu_e 
\rightarrow \nu_s}}{1-P_{\nu_e \rightarrow \nu_e}}$. In terms of mixing matrix 
elements, and considering only the 
contributions at the solar scale, one can write 
\begin{equation}
\eta_s \approx -\frac{4U_{e1}U_{e2}U_{s1}U_{s2}}{4[U_{e1}U_{e2}]^2} ~.
\end{equation}
Taking values within the allowed parameter regions for $\alpha_\mu$ and 
$\alpha_\tau$, presented in the previous section, for $m_0=0.2~eV$, we found 
that
$\eta_s \approx (1.2 - 1.9 )\times 10^{-2}$ for normal hierarchy, whereas 
$\eta_s \approx (2.7 - 3 ) \times 10^{-2}$ for inverted hierarchy. Clearly, 
this  results  
agree with the bounds given in Eq. (\ref{boundsterile}).

\begin{figure}
\begin{center}
\includegraphics[scale=0.7]{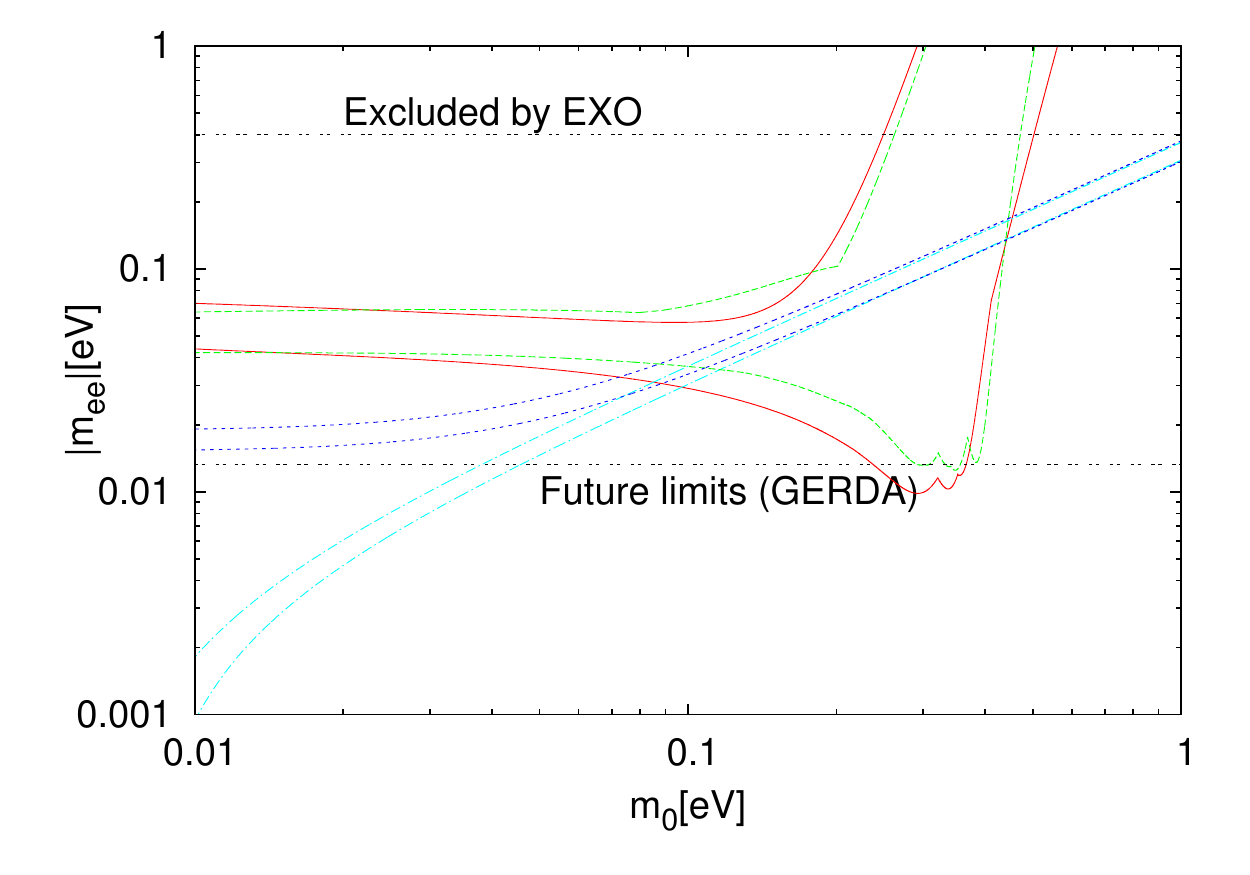}
\end{center}
\caption{ \label{ndb}
\small{ (Color online) Allowed ranges of the effective mass $|m_{ee}|$ as a 
function 
of the lightest neutrino mass $m_0$ in the case of normal (red and light blue) 
and inverted (green and blue) neutrino mass hierarchy. The lower narrowed 
regions (delimited by the light blue and blue lines) correspond to the 
conventional case of three active neutrinos. Wider upper regions (delimited by 
red and green lines) correspond to our model.}
} 
\end{figure}

On the other hand,  our sterile neutrino model will also have direct and 
interesting  implications on the effective Majorana mass term that is involved 
in neutrinoless double beta decay, now written as 
\begin{equation}
|m_{ee}|= \left | \sum_{i=1}^4 U_{ei}^2 m_i \right |.
\end{equation}
The allowed values of $|m_{ee}|$  in our model can be calculated  as a  
function of the lightest neutrino mass. Our results 
are shown in Fig.~\ref{ndb}, where, as before, we have used one sigma values 
for oscillations parameters. Consistently, sterile parameters 
were considered in the regions given as 
$0.1 \le \alpha_\mu \le 0.14$, $0.2 \le \alpha_\tau \le 0.23$ for normal 
hierarchy and $0.12 \le 
\alpha_\mu \le 0.14$, $0.24 \le \alpha_\tau \le 0.26$ for inverted hierarchy. 
This ranges are consistent with neutrino oscillation data  
in the region where $0.2\le m_0 \le 0.4$, which corresponds to the 
degenerated neutrino mass hierarchy. 

As it can be seen, the allowed parameter in our model
region is 
enhanced, compared to three-neutrino case, due to the presence of the sterile neutrino.
However,However, notice that mass hierarchy makes little difference 
for the allowed parameter space, and thus, it 
would be difficult to be experimentally identified. Nevertheless, other 
interesting differences are at hand. 
In particular, if forthcoming 
experiments were to observe a positive signal between 0.01-0.4 eV,  a 
degenerated mass spectrum with $|m_0|$ between $0.2$ and 
$0.4$ eV  might still be possible in the case of four neutrinos. 
On the other hand, the nonobservation of a signal in experiments like 
GERDA~\cite{GERDA} would practically rule out this model, which makes it 
falsifiable.

\section{Concluding remarks}

Neutrino mass models based on $\mu-\tau$ symmetry remain as an interesting 
possibility since they can provide a natural understanding for the almost 
maximal value of atmospheric neutrino mixing, and the smallness of reactor 
$\theta_{13}$ mixing, using only a couple of generic parameters that encode 
the breaking of the symmetry. 
As the analysis shows, current neutrino data are 
consistent with small values for such parameters, although, in such a scenario, 
it seems to prefer a quasidegenerate active neutrino spectrum.
As we have also pointed out, the perturbative regime of $\mu-\tau$ symmetry
breaking also provides an understanding of the, otherwise accidental,  relation 
among atmospheric and $\theta_{13}$ mixings,  which can be expressed 
through the phenomenological numerical formula 
$1/2 - \sin^2\theta_{ATM}\approx \sin\theta_{13}/\text{few}$. 
As it turns out, from our discussion, at the lower-order approximation, 
both sides of this equation are given as linear expressions in terms of the 
symmetry breaking parameters. 

On the other hand, the relative smallness of the breaking parameters can, in turn, 
be understood by the mixings of active neutrinos with a sterile neutrino, which 
by definition does not posses a flavor number, and thus neither respects 
active flavor symmetries. The model we have elaborated on in the text 
incorporates the positive features of (perturbative) $\mu-\tau$ models, allowing 
at the same time for a natural explanation of LSND/MiniBooNE 
results. As we have discussed, the model can fix all required parameters using 
oscillation neutrino observables. The allowed one sigma parameter space turns 
out to be narrow, but we consider it a nice feature of the model, since it 
allows us to explore its prediction without further approximations.
In particular, the model is
consistent with observed bounds on the sterile fraction in solar neutrino flux and 
predicts distinctive modifications on the allowed region for the neutrinoless 
double beta decay parameter, $m_{ee}$. From here, EXO limits already impose an 
upper bound for an absolute neutrino mass at about $0.5~eV$.
Moreover, even though the $m_{ee}$ region is wider compared to that 
of three neutrino scenarios, it  predicts a lower value for $m_{ee}$,   which 
would be reachable in forthcoming experiments.
As a matter of fact, our model could be excluded if no 
positive signal is found above $|m_{ee}|\sim 0.01~eV$.

Along the analysis, we have not included a Dirac \textit{CP}-violating phase. Majorana  \textit{CP} 
phases, on the other hand, have been also fixed to $0$ or $\pi$ values, which 
amount only to fixing the relative sing of the mass eigenstates. Nevertheless, it is 
interesting that our analysis shows that the only consistent combination of 
relative signs that give appropriated perturbative solutions comes when 
$m_1<0$, whereas other masses are positive. This is an intriguing result 
that could be clarified by an extended exploration of allowed  \textit{CP} phases in the 
model. Such an analysis is out of the scope of the present discussion, but it 
is part of the further work we are already undertaking.


\section*{Acknowledgments}

A.P.L. wants to thank FCFM-BUAP for the warm hospitality. 
This work was partially supported
by CONACyT, M\'exico.


\end{document}